\documentclass[aps,prd,onecolumn,showpacs,superscriptaddress,groupedaddress, nofootinbib,11pt]{revtex4}  
\usepackage{graphicx}
\usepackage{epstopdf}
\usepackage{amsmath}
\usepackage{amsfonts}
\usepackage{amssymb}
\usepackage{appendix}
\usepackage{mathtools}
\usepackage{comment}
\usepackage{bbold}
\usepackage{color}
\usepackage{slashed}
\usepackage{subfigure}
\usepackage{setspace}
\usepackage{footnote}
\usepackage[T1]{fontenc}
\usepackage{multirow}

\begin{document}

\singlespacing

\hfill NUHEP-14/03 

\title{Lepton-Flavored Dark Matter}

\author{Jennifer Kile} 
\affiliation{Institute for Fundamental Theory, Department of Physics, University of Florida, Gainesville, FL 32611, USA}

\author{Andrew Kobach} 
\affiliation{Northwestern University, Department of Physics \& Astronomy, 2145 Sheridan Road, Evanston, IL 60208, USA}

\author{Amarjit Soni}
\affiliation{Physics Department, Brookhaven National Laboratory, Upton, NY 11973 USA }

\date{\today}

\begin{abstract}
In this work, we address two paradoxes.  
The first is that the measured dark-matter relic density can be satisfied with new physics at $\mathcal{O}$(100 GeV $-$ 1 TeV), while the null results from direct-detection experiments place lower bounds of $\mathcal{O}$(10 TeV) on a new-physics scale.
The second puzzle is that the severe suppression of lepton-flavor-violating processes involving electrons, e.g.~$\mu\rightarrow 3e$, $\tau\rightarrow e \mu \mu$, etc., implies that generic new-physics contributions to lepton interactions cannot exist below $\mathcal{O}$($10-100$  TeV), whereas the $3.6\sigma$ deviation of the muon $g-2$ from the standard model can be explained by a new-physics scale $<\mathcal{O}$(1 TeV).   
Here, we suggest that it may not be a coincidence that both the muon $g-2$ and the relic density can be satisfied by a new-physics scale $\lesssim 1$ TeV. 
We consider the possibility of a gauged lepton-flavor interaction that couples at tree level only to $\mu$- and $\tau$-flavored leptons and the dark sector.  Dark matter thus interacts appreciably only with particles of $\mu$ and $\tau$ flavor at tree level and has loop-suppressed couplings to quarks and electrons.
Remarkably, if such a gauged flavor interaction exists at a scale $\mathcal{O}$(100 GeV $-$ 1 TeV), it allows for a consistent phenomenological framework, compatible with the muon $g-2$, the relic density, direct detection, indirect detection, charged-lepton decays, neutrino trident production, and results from hadron and $e^+e^-$ colliders.  
We suggest experimental tests for these ideas at colliders and for low-energy observables.  
\end{abstract}

\pacs{95.35.+d, 11.30.Hv, 14.60.-z}
\maketitle


\section{Introduction}

In this work, we attempt to address two ongoing puzzles in particle physics. The first is that if dark matter is a thermal relic, its annihilation cross section requires new physics at the electroweak scale, i.e., in the range $\mathcal{O}$(100 GeV $-$ 1 TeV). However, the null results from direct-detection experiments constrain that a new-physics scale between dark matter and nucleons must be $> \mathcal{O}$(10 TeV) for a dark-matter mass $\gtrsim 10$ GeV. This tension between the relic density and direct detection may be pointing to the possibility that dark matter does not  couple to quarks at tree level.   Rather, dark matter may couple primarily via interactions at the electroweak scale to other particles in the standard model, e.g., leptons.  If so, such a dark-matter candidate can satisfy the measured relic density at tree level and accommodate the null results from direct detection by giving rise to interactions between dark matter and quarks at the loop level.

If interactions at the electroweak scale exist between dark matter and leptons, then one generically expects such interactions between leptons themselves.  There may be evidence for such an interaction given that the current $3.6\sigma$ deviation of the muon $g-2$ from the standard model value could be explained by new interactions at a scale $<\mathcal{O}$(1 TeV).   
However, the possible existence of new physics at this scale introduces a second puzzle:~interactions at such a scale do not manifest themselves via other processes among charged leptons. For example, flavor-violating processes such as $\mu\rightarrow 3e$, $\tau\rightarrow e\mu\mu$, $\tau\rightarrow ee\mu$, $\tau\rightarrow 3e$, $\mu\rightarrow e\gamma$~\cite{Agashe:2014kda} constrain new-physics scales to be $> \mathcal{O}$(10$-$100 TeV). Additionally, flavor-conserving processes, such as lepton production at LEP, constrain new physics to scales above several TeV~\cite{ALEPH:2005ab, Schael:2013ita,Abbiendi:1999wm}.  Some of the strong experimental constraints on interactions among leptons can be found in Table~\ref{table:constraints}.  
Since most of these strong constraints come from processes that involve electrons, this motivates
the possibility of a new leptonic interaction at the electroweak scale under which electrons, like quarks, do not effectively participate.

\begin{table}[ht]
\centering
\begin{tabular}{| c | c |}
\hline
Observable & Limit \\ \hline \hline
Br($\mu\rightarrow 3e$) & $<1.0\times 10^{-12}$~\cite{Agashe:2014kda} \\ \hline
Br($\mu\rightarrow e\gamma$) & < $5.7\times10^{-13}$~\cite{Agashe:2014kda} \\ \hline
Br($\tau\rightarrow 3e$) & < $2.7\times 10^{-8}$~\cite{Agashe:2014kda} \\
Br($\tau\rightarrow e^-\mu^+\mu^-$) & < $2.7\times 10^{-8}$~\cite{Agashe:2014kda} \\
Br($\tau\rightarrow e^+\mu^-\mu^-$) & < $1.7\times 10^{-8}$~\cite{Agashe:2014kda} \\
Br($\tau\rightarrow \mu^-e^+e^-$) & < $1.8\times 10^{-8}$~\cite{Agashe:2014kda}\\
Br($\tau\rightarrow \mu^+e^-e^-$) & < $1.5\times 10^{-8}$~\cite{Agashe:2014kda} \\
Br($\tau\rightarrow 3\mu$) & < $2.1\times 10^{-8}$~\cite{Agashe:2014kda} \\ \hline
Br($\tau\rightarrow \mu\gamma$) & < $4.4\times 10^{-8}$~\cite{Agashe:2014kda} \\ 
Br($\tau\rightarrow e\gamma$) & < $3.3\times 10^{-8}$~\cite{Agashe:2014kda} \\ \hline
$\mu-e$ conversion & $\Lambda \gtrsim 10^3$ TeV~\cite{deGouvea:2013zba} \\ \hline 
$e^+e^- \rightarrow e^+ e^-$ & $\Lambda \gtrsim 5$ TeV~\cite{Schael:2013ita} \\ 
$e^+e^- \rightarrow \mu^+ \mu^-$ & $\Lambda \gtrsim 5$ TeV~\cite{Schael:2013ita} \\ 
$e^+e^- \rightarrow \tau^+ \tau^-$ & $\Lambda \gtrsim 4$ TeV~\cite{Schael:2013ita} \\ \hline
\end{tabular}
\caption{Constraints on lepton-flavor violating and conserving processes.  For the last four observables, the experimental null results are given in terms of a dimension-6 operator, suppressed by two orders of $\Lambda$, which can be interpreted as the nominal scale of new physics. }
\label{table:constraints}
\end{table}

Here, we consider the possibility that the occurrence of these two paradoxes is not a coincidence.  We find that they can be resolved simultaneously if we consider a gauged lepton-flavor interaction at the electroweak scale under which both leptons and dark matter are charged.  We call this framework ``lepton-flavored dark matter'' (LFDM).  Here, we consider the simplified case where the interaction only involves $\mu$- and $\tau$-flavored leptons and dark matter; thus, dark matter interacts only with particles of $\mu$ and $\tau$ flavor at tree level. We perform a model-independent analysis of this scenario and find that it can lead to a consistent framework where the relic density, direct detection,
indirect detection, results from hadron and $e^+ e^-$ colliders,  and low-energy measurements are compatible with flavor gauge bosons with electroweak-scale masses, i.e., $\mathcal{O}$(100 GeV$ - $1 TeV).

Many have investigated the idea that dark matter does not interact with quarks at the tree level.  Most of these analyses  assume an interaction between dark matter and leptons that does not distinguish between lepton flavors~\cite{Chang:2014tea, Schmidt:2012yg, Agrawal:2011ze, Carone:2011iw, Ko:2010at, Haba:2010ag, Farzan:2010mr, Chun:2009zx, Cohen:2009fz, Davoudiasl:2009dg, Ibarra:2009bm, Kyae:2009jt, Chen:2008dh, Baltz:2002we, Bai:2014osa, Schwaller:2013hqa, Basso:2012ti, Carone:2011ur, Chao:2010mp, Khalil:2009nb, Cao:2009yy, Freitas:2014jla}, although a few have considered more general analyses of gauged flavor interactions~\cite{Das:2013jca, Fox:2008kb, Bi:2009uj, Kopp:2014tsa, Hamze:2014wca, Bell:2014tta, Lee:2014rba, Baek:2008nz}.  The LFDM framework allows for a more general analysis of interactions that  involve only dark matter and leptons at the tree level; it permits different coupling strengths between lepton flavors, off-diagonal flavor couplings, and lepton-flavor violation.  For a review of flavored dark matter, see Ref.~\cite{Kile:2013ola} and the references therein.

Our analysis is outlined as follows.  In Section~\ref{interactions}, we introduce a parameterization of LFDM that, to a good approximation, can encapsulate most models of flavor-conserving LFDM that involve only $\mu$- and $\tau$-flavored leptons.  In Section~\ref{observ}, we survey the relevant constraints on LFDM from the relic density, direct-detection, the muon $g-2$, indirect detection, and high-energy observables at LEP and the LHC. In Section~\ref{scans}, we explore the important constraints via a parameter scan.  We discuss the possibility of lepton-flavor violation (LFV) in Section~\ref{LFV}, and  in Section~\ref{conclusion}, we offer conclusions from our analysis and prospects for future experimental and theoretical investigations.

\section{Lepton-Flavor Interactions}
\label{interactions}

In this section, we lay out the general framework that we use to analyze LFDM.  We take dark-sector particles to be comprised of Dirac fermions, sharing common gauged lepton-flavor interactions along with right- and left-handed charged leptons.  We assume that neutrinos have Dirac masses and consequently introduce three right-handed partners also charged under the lepton-flavor symmetry.\footnote{We do not address the possibility that introducing additional light degrees of freedom can effect measurements of $n_\text{eff}$ from the cosmic-microwave background.}
We take the strong constraints on processes that include electrons as evidence that particles with electron flavor do not effectively participate in lepton-flavor interactions at the electroweak scale.

If the muon $g-2$ anomaly is due to LFDM, this may be an indication that the interactions are purely vector-like.\footnote{If the flavor interactions were purely left- or right-handed, they cannot account for the muon $g-2$ anomaly, since they would decrease the value of $a_\mu$.}  It is possible, however, that due to different rotations between mass and interaction eigenstates in the right- and left-handed sectors, differences in left- and right-handed couplings can arise in the mass eigenstate basis.  This can introduce interactions that deviate  from purely vector interactions. Since flavor-violating couplings among charged leptons can be strongly constrained, we assume that the LFDM interactions do not mediate LFV among charged leptons at the tree level; we thus take the mass and flavor interaction bases in the charged-lepton sector to be closely aligned and take the deviations from purely-vector interactions in the mass basis to be negligibly small.  
However, because we have essentially no constraints on the flavor structure of the dark sector, we allow different lepton-flavor couplings between left- and right-handed components of dark matter.  We return to the topic of LFV in Section~\ref{LFV}.

We create a phenomenological Lagrangian with dimension-four ($d=4$) operators that can encapsulate most models of LFDM that do not violate lepton flavor at the tree-level, but do permit off-diagonal lepton-flavor vertices. One can introduce an arbitrary number of gauge bosons, categorized by whether they couple to diagonal or off-diagonal lepton-flavor currents.  We denote their mass eigenstates by $X$ and $Y$, respectively.  
We presently consider that there are two $X$ bosons and one $Y$ boson, which allows for a sufficient number of terms to account for the phenomenology associated with a wide variety of specific models.\footnote{Two $X$ bosons allow us to independently vary the strength of the flavor-conserving four-Fermi interactions between $\chi$ and the charged leptons, and among the charged leptons themselves.  If only a single $X$ boson were used, for example, the four-$\mu$, four-$\tau$, and $\bar{\mu}\mu\bar{\tau}\tau$ effective interactions would have related coefficients.  A second $X$ boson allows for these coefficients to be independent.}  
There can also be an arbitrary number of dark-sector species that participate in the lepton-flavor interactions. The relic density and direct detection are only sensitive to the lightest dark-sector state, which we call $\chi$.

We parameterize these lepton-flavor interactions with the following phenomenological Lagrangian:\footnote{We note that achieving a Lagrangian like in Eq.~(\ref{dim4noLFV1DM}) in a general model of flavor is non-trivial; care must be taken to not have flavor violation in conflict with experimental constraints.  Our motivation in choosing the form of Eq.~(\ref{dim4noLFV1DM}) is thus phenomenological, and not derived from a particular model.}

\begin{eqnarray}
\label{dim4noLFV1DM}
\mathcal{L} &\supset& \displaystyle\sum_{i=1,2} X_{i\alpha} \Big[ k_{i\mu\mu} J^\alpha_\mu + k_{i\tau\tau} J^\alpha_\tau + k'_{iL} \overline{\chi}_L \gamma^\alpha \chi_L + k'_{iR} \overline{\chi}_R \gamma^\alpha \chi_R \Big]  \nonumber  \\
&&\quad +~  Y_\alpha \Big[ h_{\mu\tau} K^\alpha + h'_L \overline{\chi}_L \gamma^\alpha \chi_L + h'_R \overline{\chi}_R \gamma^\alpha \chi_R \Big]  \\
&&\quad +~ Y_\alpha^\dagger \Big[ h_{\mu\tau} (K^\alpha)^\dagger + h'_L \overline{\chi}_L \gamma^\alpha \chi_L + h'_R \overline{\chi}_R \gamma^\alpha \chi_R \Big], \nonumber
\end{eqnarray}
where
\begin{eqnarray}
J^\alpha_\mu &=&  \overline{\mu}_L \gamma^\alpha \mu_L +  \overline{\mu}_R \gamma^\alpha \mu_R  + \overline{\nu}_{L\mu}\gamma^\alpha \nu_{L\mu} + \overline{\nu}_{R\mu}\gamma^\alpha \nu_{R\mu}, \\
J^\alpha_\tau &=& \overline{\tau}_L \gamma^\alpha \tau_L +  \overline{\tau}_R \gamma^\alpha \tau_R  + \overline{\nu}_{L\tau}\gamma^\alpha \nu_{L\tau} + \overline{\nu}_{R\tau}\gamma^\alpha \nu_{R\tau}, \\
K^\alpha &=&  \overline{\mu}_L \gamma^\alpha \tau_L +  \overline{\mu}_R \gamma^\alpha \tau_R   + \overline{\nu}_{L\mu}\gamma^\alpha \nu_{L\tau} + \overline{\nu}_{R\mu}\gamma^\alpha \nu_{R\tau}.
\end{eqnarray}
We take $\nu_{R\mu}$ and $\nu_{R\tau}$ to be defined by their interactions under the flavor group; while the interaction basis does not coincide with the mass basis in the neutrino sector, our analysis is not sensitive to this mixing.
The differences between the couplings $k$ and $k'$ (and between $h$ and $h'$) account for both the possibility that the charged leptons and the dark sector are not in the same representation of the flavor group and the possible occurrence of non-negligible mixing in the dark sector.\footnote{In general, the dark-matter coupling $h'$ can be complex due to the presence of CP-violating phases when rotating from the flavor basis to the mass basis.  However, since our analysis is not sensitive to the effects of CP violation, we assume $h'$ is real for simplicity.}
We neglect a tree-level kinetic mixing operator between $X$ and the hypercharge gauge boson, $X_{\mu\nu}B^{\mu\nu}$, which can permit the lepton-flavor interactions to couple to quarks and electrons at tree level.  If the flavor gauge symmetry is not a $U(1)$, this term is disallowed; in the case of a $U(1)$ symmetry, we expect this choice to be conservative for the constraints relevant for this analysis.

\section{Flavor-Conserving Observables}
\label{observ}

\subsection{Direct Detection and Relic Density}
\label{directdetection}


A basic requirement for any dark-matter candidate is that it has a relic density not in conflict with observation.   If dark matter consists of a single thermal relic, it implies that Dirac dark matter has a thermally-averaged annihilation cross section to SM particles of $\left< \sigma v \right> \approx 4.4\times10^{-26}$ cm$^3$/s~\cite{Steigman:2012nb}.  The dark matter will annihilate to pairs of charged leptons or neutrinos.\footnote{We assume that $M_{X_1}, M_{X_2}, M_Y > m_\chi$  and thus neglect the possibility of dark-matter annihilations to the flavor gauge bosons.  We check the self-consistency of this assumption in Section~\ref{scans}.}  If the masses of the final-state leptons are negligible, the non-relativistic annihilation cross section is the following, according to the parameterization of LFDM given in Eq.~(\ref{dim4noLFV1DM}), 
\begin{equation}
\label{relic}
\left< \sigma v \right> =  \frac{m_\chi^2}{2\pi} \left\{  \frac{2h_{\mu\tau}^2 ( h'_L + h'_R)^2}{M^4_Y} + \left[ \left( \displaystyle\sum_{i=1,2}\frac{k_{i\mu\mu}(k'_{iL}+k'_{iR})}{M_{X_i}^2}   \right)^2  + \left( \sum_{i=1,2}\frac{k_{i\tau\tau}(k'_{iL}+k'_{iR})}{M_{X_i}^2}   \right)^2 \right]  \right\}.
\end{equation}
If $\left< \sigma v \right> = m_\chi^2/\Lambda^4 \sim 4.4\times10^{-26}$ cm$^3$/s, where $\Lambda$ is the effective scale of the interaction, then $\Lambda \sim  (130\text{ GeV})\sqrt{m_\chi/\text{GeV}}$.  If LFDM is allowed to be only a subset of dark matter, then smaller scales can be obtained.

\begin{figure}[tbp]
\begin{center}
\includegraphics[scale=0.55]{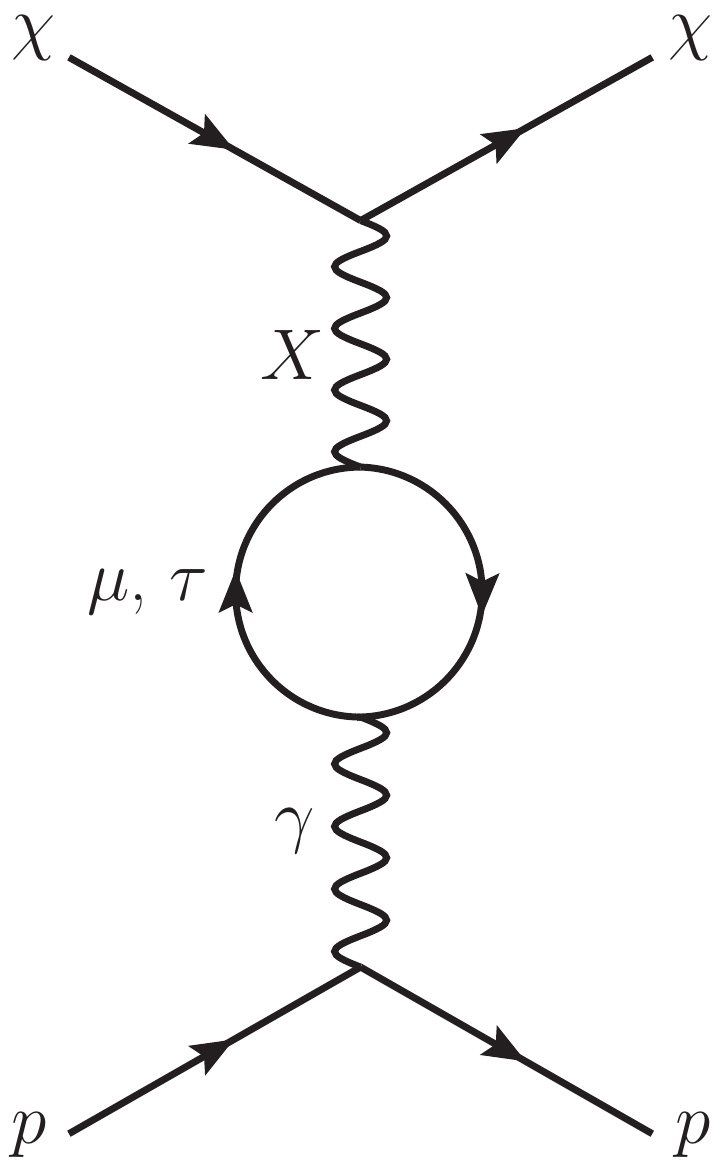}
\caption{The 1-loop diagram that contributes to direct detection by mediating the interaction between lepton-flavored dark matter and protons, as parameterized in Eq.~(\ref{dim4noLFV1DM}). }
\label{dd}
\end{center}
\end{figure}

LFDM does not interact with quarks at the tree level; the direct-detection cross section first occurs at 1-loop, as shown in Fig.~\ref{dd}.   Because the photon only couples to charged particles with flavor-diagonal couplings, the $Y$ boson does not participate in this process.  
We estimate the value of this diagram as the running between the energy scale relevant for direct detection, which we take to be the mass of the lepton in the loop, and a renormalization scale $\mu$, which we assume is of order the mass of the flavor gauge bosons, $\mu\sim M_{X_i}$ (see also Refs.~\cite{Kopp:2009et, Fox:2011fx}).  We obtain 
\begin{equation}
\label{ddxsect}
\sigma(\chi p \rightarrow \chi p) = \frac{\alpha^2\mu_N^2}{36\pi^3}   \left[ \displaystyle\sum_{i=1,2} \frac{k_{i\mu\mu}(k'_{iR} +k'_{iL})}{M_{X_i}^2} \ln\left(\frac{\mu^2}{m_\mu^2}\right) + \frac{k_{i\tau\tau} (k'_{iR} +k'_{iL}) }{M_{X_i}^2} \ln\left(\frac{\mu^2}{m_\tau^2}\right)  \right]^2 ,
\end{equation}
where $\mu_N=m_\chi m_p/(m_\chi + m_p)$.

The LUX experiment places limits on the spin-independent interaction cross section between dark matter and nucleons to be $\lesssim \mathcal{O}$($10^{-45} - 10^{-44}$ cm$^2$) for $m_\chi > 10$ GeV~\cite{Akerib:2013tjd}.  In order to interpret these results as limits on the interaction cross section between dark matter and protons, one can scale the experimental results by a factor of $Z^2/A^2$, where $Z$ and $A$ are the atomic number and atomic mass number, respectively, of the detector's interaction material. 
If $\sigma(\chi p \rightarrow \chi p) \sim \alpha^2 m_p^2/(36\pi^3\Lambda^4)$, a cross section upper limit of $10^{-45}$ cm$^2$ at LUX would suggest that $\Lambda \gtrsim $ 400 GeV, which is consistent with electroweak-scale LFDM gauge boson masses.

\subsection{Low-Energy Observables}
\label{g-2}

\subsubsection{Muon $g-2$} 

\begin{figure}[htbp]
\begin{center}
\subfigure[]{\label{g-21}\includegraphics[width=0.3\textwidth]{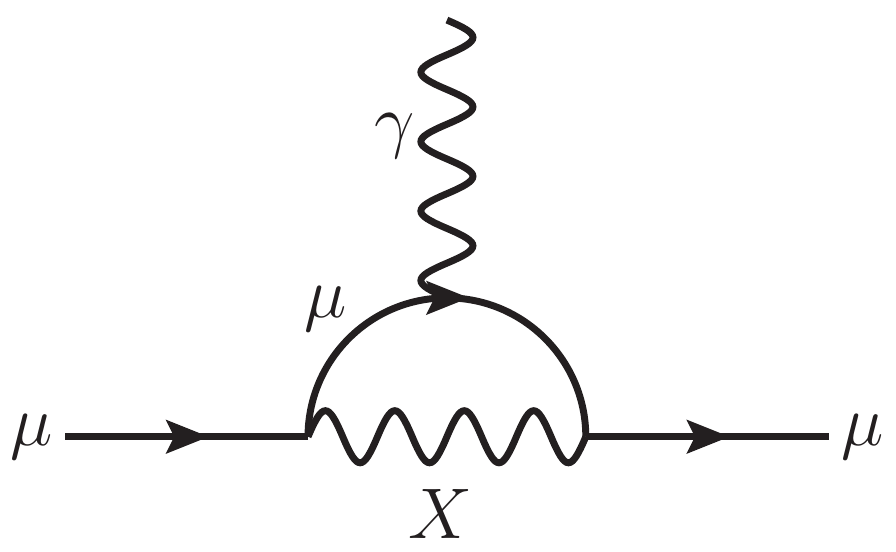}}
\hspace{0.5in}
\subfigure[]{\label{g-22}\includegraphics[width=0.3\textwidth]{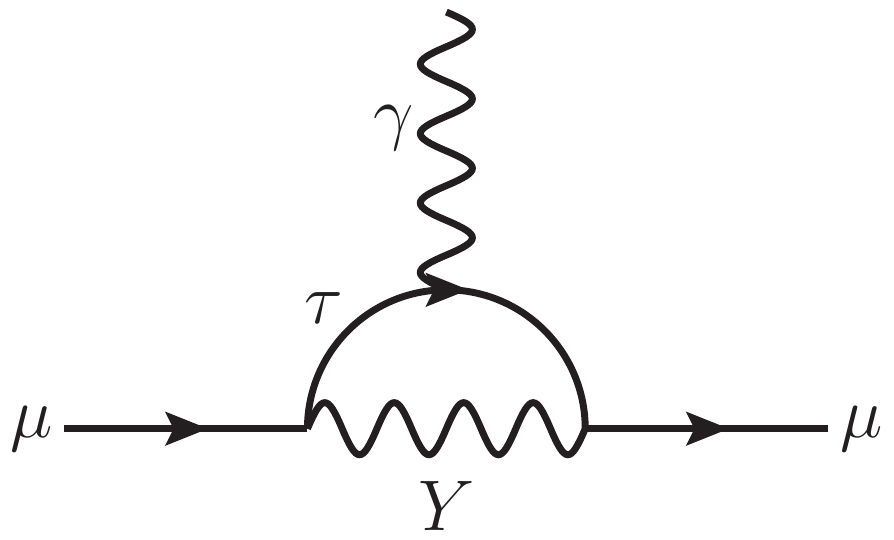}}
\caption{The 1-loop contribution to the muon $g-2$ due to the lepton-flavor interactions, according to the parameterization in Eq.~(\ref{dim4noLFV1DM}).  }
\label{mag}
\end{center}
\end{figure}
Currently, the only low-energy measurement that deviates significantly from the SM is the muon $g-2$, whose measured value is 3.6$\sigma$ larger than the SM expectation, $a_\mu^\text{SM}$, i.e., $\delta_{a_\mu} = a_\mu^\text{exp} - a_\mu^\text{SM} = (28.8 \pm 8.0)\times 10^{-10}$~\cite{Agashe:2014kda}.  This discrepancy can be accounted for by LFDM through the diagrams shown in Fig.~\ref{mag}.  Here, the loop contains either a muon and an $X$ boson, or a tau and a $Y$ boson.  The value of $a_\mu$ is given by  
\begin{equation}
a_\mu = a_\mu^{SM} + \frac{h_{\mu\tau}^2}{4\pi^2} \frac{m_{\mu}^2}{M_{Y}^2} \left( \frac{m_\tau}{m_\mu} - \frac{2}{3} \right) +  \displaystyle\sum_{i=1,2}\frac{k_{i\mu\mu}^2}{12\pi^2} \frac{m_{\mu}^2}{M_{X_i}^2} . 
\end{equation}
If the theoretical prediction were to be brought to a value within 95\% CL of the experimental value,\footnote{For the muon $g-2$, and throughout this analysis, we use a $\chi^2$ function to estimate limits for different CL's. } i.e., $1.3\times10^{-9}<\delta_{a_\mu}<4.4\times10^{-9}$, then $1/(270\text{ GeV})^2 <\sum_{i}k^2_{i\mu\mu}/M_{X_i}^2 < 1/(140\text{ GeV})^2$ for the case where only muons run in the loop, and $1/(1.9\text{ TeV})^2 <h^2_{\mu\tau}/M_Y^2 < 1/(1.0\text{ TeV})^2$ when only taus run in the loop. These values are consistent with the requirements from the dark-matter relic density and direct detection.

\subsubsection{Tau Decays}
\label{taudecays}
Lepton-flavored interactions mediated by $Y$ bosons will constructively interfere with the SM lepton decays of the tau lepton, $\tau^- \rightarrow \mu^-  \nu_\tau\overline{\nu}_\mu$.  Using the parameterization in Eq.~(\ref{dim4noLFV1DM}), the width for such decays is 
\begin{equation}
\Gamma(\tau^- \rightarrow \mu^-  \nu_\tau\overline{\nu}_\mu) \simeq \frac{m_\tau^5}{\pi^3} \left(\frac{G_F^2}{192} + \frac{\sqrt{2} G_F h^2_{\mu\tau}}{384~ M_{Y}^2}  \right) \left( 1- \frac{8m_\mu^2}{m_\tau^2}  \right) , 
\end{equation}
where $G_F$ is the Fermi constant.  In the SM, the ratio of $\Gamma(\tau^- \rightarrow \mu^-  \nu_\tau\overline{\nu}_\mu) / \Gamma(\tau^-\rightarrow e^-\nu_\tau\overline{\nu}_e)$ is estimated to be $0.9726$, and this ratio has been measured to be $0.979 \pm 0.004$~\cite{Agashe:2014kda}.  
This measurement suggests that the 95\% CL band requires $ h_{\mu\tau}^2 / M_Y^2  \lesssim 1/(2.0 \text{ TeV})^2 $.  This poses a mild tension with the muon $g-2$ if only the $Y$ boson participates in LFDM.

\subsection{Neutrino Trident Production}
\label{trident}

\begin{figure}[tbp]
\begin{center}
\includegraphics[scale=0.65]{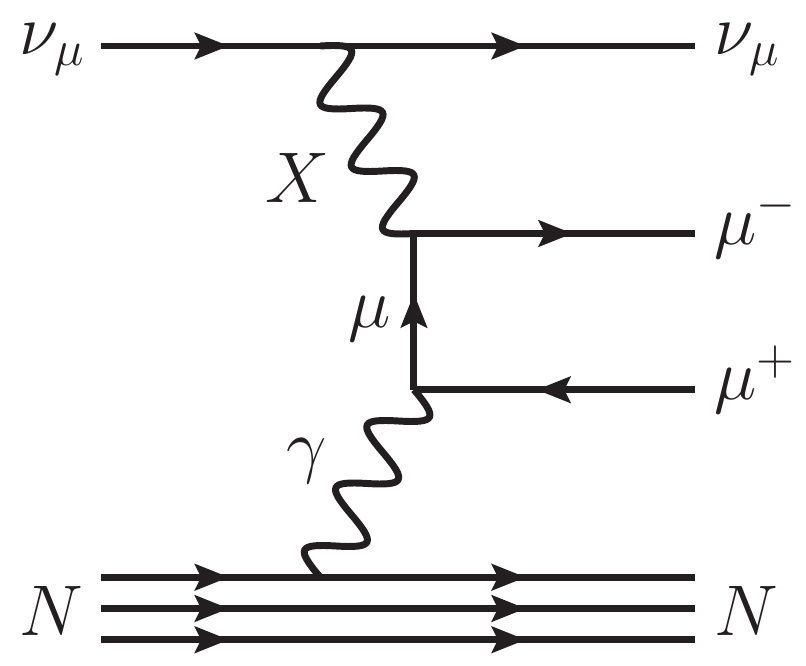}
\caption{The contribution to neutrino trident production from $X$ bosons, according to the parameterization in Eq.~(\ref{dim4noLFV1DM}). }
\label{tridentfig}
\end{center}
\end{figure}

Interactions of the $X$ flavor gauge bosons can contribute to measurements of neutrino trident production, as depicted in Fig.~\ref{tridentfig}.  As calculated by the authors of Refs.~\cite{Altmannshofer:2014cfa, Altmannshofer:2014pba}, an $X$ boson can constructively interfere with the SM, resulting in an increase in the measured cross section compared to that of the SM expectation:
\begin{equation}
\label{tridentequ}
\frac{\sigma_\text{SM+X}}{\sigma_\text{SM}} = \frac{1+\left(1+4\sin^2\theta_w +\displaystyle \frac{\sqrt{2}}{G_F} \sum_i \frac{k^2_{i\mu\mu}}{M_{X_i}^2} \right)^2}{1+\left(1+4\sin^2\theta_w\right)^2}.
\end{equation}

The results of the CHARM-II~\cite{Geiregat:1990gz} and CCFR~\cite{Mishra:1991bv} experiments are $\sigma_\text{data}/\sigma_\text{SM} = 1.58\pm 0.57$ and $0.82 \pm 0.28$, respectively.  Together, these data suggest that, at 95\% CL, 
\begin{equation}
\displaystyle\sum_i \frac{k_{i\mu\mu}^2}{M_{X_i}^2} < \frac{1}{(490 \text{ GeV})^2}. 
\end{equation}
This constraint from neutrino trident production strongly disfavors the hypothesis that the value of the muon $g-2$ receives a contribution from $X$ bosons alone.  Likewise, as mentioned in Section~\ref{taudecays}, a parameterization of LFDM with only $Y$ bosons is constrained by tau decays and likewise cannot account for the measured value of the muon $g-2$.  However, if one permits both $X$ and $Y$ bosons to contribute to the muon $g-2$, as shown in Fig.~\ref{mag}, then one can account for the muon $g-2$. We will explore these ramifications for direct detection and the LHC in Section~\ref{scans}.

\subsection{Indirect Detection}
\label{indirectdetection}
Other possible constraints on LFDM come from indirect detection. Our scenario is not relevant for explaining the excesses in the positron fraction from AMS \cite{Aguilar:2013qda}, Pamela \cite{Adriani:2013uda}, and Fermi-LAT \cite{FermiLAT:2011ab}, which require an annihilation cross section to leptons approximately two to three orders of magnitude larger than that expected for a thermal relic \cite{Jin:2013nta,Cirelli:2008pk,Yuan:2013eja}.  
Since we assume that $\chi$ annihilates half the time to neutrinos, indirect detection is only relevant for our scenario if it constrains the annihilation cross-section to $\mu^+\mu^-$ or $\tau^+\tau^-$ to be less than half of the thermal relic value.  For values of the dark matter mass $m_{\chi}\gtrsim 100$ GeV, indirect detection bounds are not sufficiently stringent to rule out the thermal cross section. 

However, for somewhat light dark matter, $m_{\chi} \sim \mbox{few}\times 10$ GeV, indirect-detection constraints may be important, but are subject to significant uncertainties.  Dark-matter annihilation to $\mu^+\mu^-$ can be constrained using the AMS positron fraction to be below the thermal relic value for dark-matter masses below $100$ GeV~\cite{Bergstrom:2013jra}, but systematic uncertainties can alter the limits on the annihilation cross-section by a factor of a few.  (For results using the positron flux, also see Ref.~\cite{Ibarra:2013zia}.)

Indirect-detection constraints are important, however, for the low-mass region $m_{\chi}\lesssim 10\mbox{ GeV}$.  The annihilation of light dark matter can be constrained by the effects of its resultant energy injection on the CMB; thermal relics with masses below several GeV are disfavored \cite{Madhavacheril:2013cna,Hutsi:2011vx,Cirelli:2009bb}.  Annihilations of light dark matter into $\tau^+\tau^-$ are constrained by Fermi-LAT observations to be as small as a few $\times 10^{-27}$ cm$^3$/s~\cite{Hooper:2012sr,Gordon:2013vta,Tavakoli:2013zva,fortheFermiLAT:2013naa}.  Fermi-LAT also gives similar constraints, ${\cal O}(10^{-27}-10^{-26}$ cm$^3$/s), on annihilations of light dark matter into $\mu^+\mu^-$ \cite{Berlin:2013dva}.  Slightly tighter constraints are also possible from AMS, but are subject to effects of solar modulation for dark-matter masses $m_{\chi}< {\cal O}(15\mbox{ GeV})$ \cite{Tavakoli:2013zva,Ibarra:2013zia,Bergstrom:2013jra}. 

We note, however, that these constraints all give limits on the annihilation cross section of light dark matter to leptons which are smaller than that of a thermal relic by a factor of order unity.  While we also include annihilations of dark matter to neutrinos (equal in magnitude to dark matter annihilations to charged leptons), some mild tension remains between bounds from indirect detection and the thermal cross section.  Interestingly, some works have considered possible evidence for dark matter annihilation to leptons from indirect detection with cross sections slightly below those of a thermal relic \cite{Hooper:2012ft,Gordon:2013vta,Huang:2013pda,Hooper:2013rwa,Daylan:2014rsa,Berlin:2013dva,Hooper:2010im,Belikov:2011pu}.  We will briefly return to the subject of indirect-detection constraints when we discuss the parameter scans in Section~\ref{scans}.

\subsection{High-Energy Colliders}

\label{HEobservables}

\begin{figure}[htbp]
\begin{center}
\subfigure[]{\label{zdecay1}\includegraphics[width=0.25\textwidth]{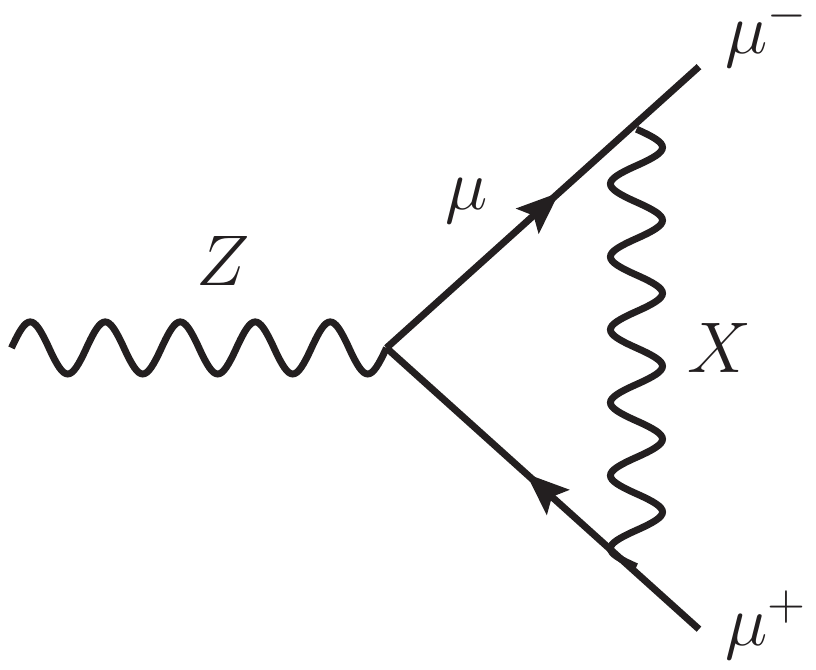}}
\hspace{0.5in}
\subfigure[]{\label{zdecay2}\includegraphics[width=0.25\textwidth]{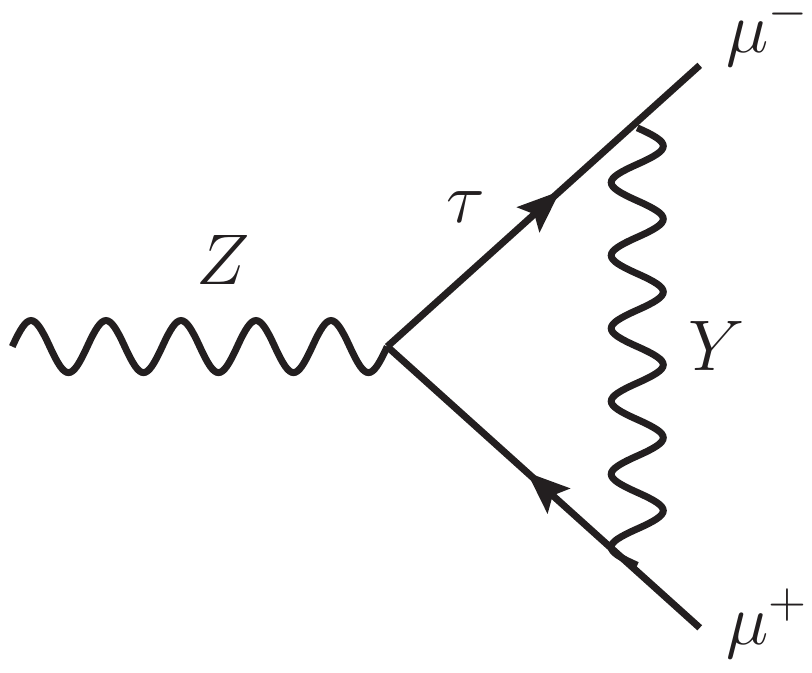}}
\subfigure[]{\label{LEPLHC}\includegraphics[width=0.41\textwidth]{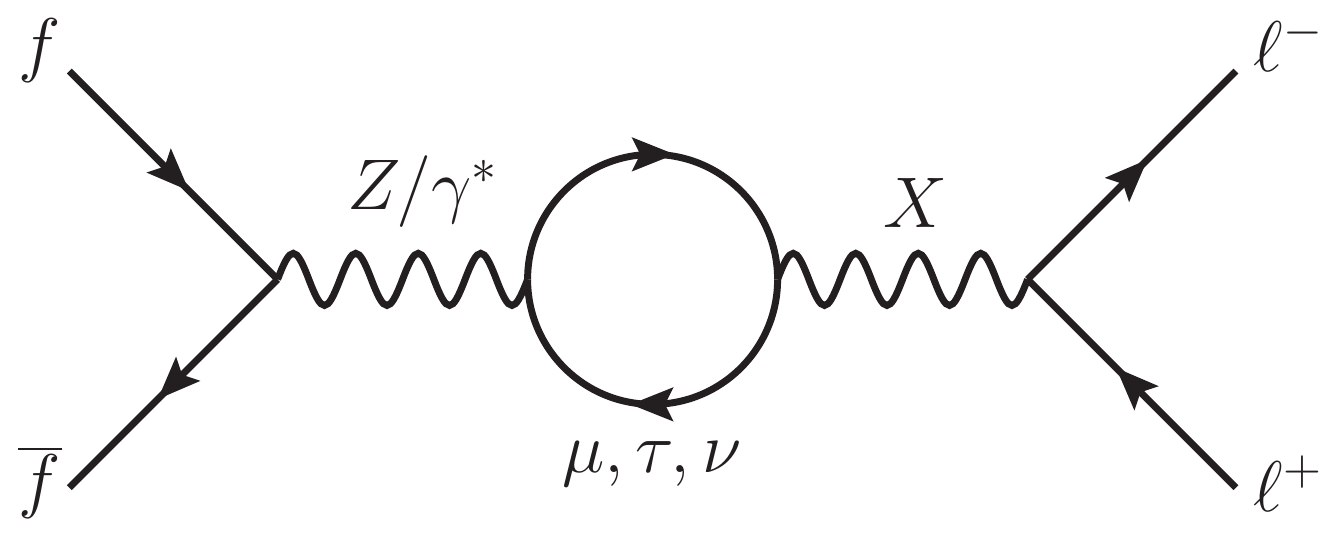}}
\caption{(a) and (b): The 1-loop correction to the $Z$-decay vertex due to flavor gauge bosons, $X$ and $Y$.  (c): The 1-loop diagram for kinetic mixing  between $X$ and the hypercharge gauge boson $B$ at hadron and lepton colliders (where $f\bar{f}$ can be $q\overline{q}$ or $e^+e^-$), which can contribute to the $Z$-lepton couplings, resonant $X$ production, and effective four-fermion operators. }
\label{collider}
\end{center}
\end{figure}

As discussed above, the requirements from the muon $g-2$ and the relic density imply that the preferred scale of LFDM is the electroweak scale, i.e., $\mathcal{O}$(100 GeV$-$1 TeV), which is within the reach of collider experiments. While the lepton-flavor gauge bosons do not effectively couple to electrons and quarks at tree level, they can contribute, however, to the processes $e^+e^-\rightarrow \mu^+\mu^-, \tau^+\tau^-$ and $q\bar{q}\rightarrow \mu^+\mu^-, \tau^+\tau^-$ via the one-loop diagrams in Fig.~\ref{collider}.  Potentially relevant measurements include constraints on resonance searches, effective operators, and couplings of the $Z$ boson to $\mu^+\mu^-$ and $\tau^+\tau^-$ at LEP, as well as resonance searches at LHC.

\subsubsection{LEP}

The diagrams in Fig.~\ref{collider} can potentially give new physics contributions to three measurements from the LEP experiments. First, these diagrams may affect the limits from LEP on effective operators of the form $\Lambda^{-2}(\overline{e}_i\gamma^\alpha e_i)(\overline{\ell}_j\gamma_\alpha \ell_j)$, where $\ell=\mu, \tau$, and $i,j=R$ or $L$, for which $\Lambda$ is constrained to be larger than several TeV~\cite{Schael:2013ita}.  Second, lepton-flavor physics can produce corrections to the $Z$-lepton couplings, as depicted in the three diagrams in Fig.~\ref{collider}. (The diagram in Fig.~\ref{LEPLHC} only contributes to the $Z$-lepton couplings if the $s$-channel SM gauge boson is a $Z$.)  Lastly, narrow resonance searches at LEP can constrain LFDM via the diagram in Fig.~\ref{LEPLHC}.  Since all of the diagrams in Fig.~\ref{collider} are loop-suppressed, the limits on the scale of new physics resulting from these considerations are typically weak, $M_X \gtrsim 200\mbox{ GeV}$.  However, we give the case of light $X$ bosons special consideration, because this is similar to the scale of LFDM interactions suggested by $g-2$ for the case where only $X$ bosons and muons contribute.  Here, we do not consider the contribution from the diagram in Fig.~\ref{zdecay2}, since, compared to the other diagrams in Fig.~\ref{collider}, it has a negligible effect on collider observables because of the low-energy results in Section~\ref{g-2}. 

First, we note that when the diagram in Fig.~\ref{LEPLHC} contains an $s$-channel photon, its contribution to $e^+e^-\rightarrow \mu^+\mu^-, \tau^+\tau^-$ is non-resonant at LEPII center-of-mass energies between 130 GeV and 209 GeV.  We obtain a rough constraint on this contribution by approximating it as an effective operator $\Lambda^{-2}(\overline{e}\gamma^\alpha e)(\overline{\ell}\gamma_\alpha \ell)$, which should be approximately valid for values of $M_X \gtrsim 250$ GeV.  For the case of constructive interference with the SM, the 95\% CL lower limit on $\Lambda$ for this operator is $5.3$ TeV for $\ell=\mu$ and $4.5$ TeV for $\ell=\tau$, and for the case of destructive interference, the limits are $4.6$ TeV and $3.9$ TeV, respectively~\cite{Schael:2013ita}.   We obtain 
\begin{align}
\label{eq:eemmttop1}
\frac{-1}{(4.6\mbox{ TeV})^2}< \frac{e^2}{12\pi^2 }  \left[ \frac{5}{3}+ \ln\left(\frac{\mu^2}{q^2}\right)\right]  \sum_{i} \frac{k_{i\mu\mu}(k_{i\mu\mu}+k_{i\tau\tau})}{ M_{X_i}^2} <\frac{1}{(5.3\mbox{ TeV})^2},\\
\label{eq:eemmttop2}
\frac{-1}{(3.9\mbox{ TeV})^2}<  \frac{e^2}{12\pi^2 }  \left[ \frac{5}{3}+ \ln\left(\frac{\mu^2}{q^2}\right)\right]  \sum_{i} \frac{k_{i\tau\tau}(k_{i\mu\mu}+k_{i\tau\tau})}{ M_{X_i}^2}  <\frac{1}{(4.5\mbox{ TeV})^2},
\end{align}
where $q^2$ is the center-of-mass energy.  While the results in Ref.~\cite{Schael:2013ita} are derived from all LEPII energies, we take the approximation\footnote{Here, we choose $q^2=(189 \text{ GeV})^2$ because it is the lowest LEPII center-of-mass energy with a large luminosity $(>100 \text{ pb}^{-1})$.  Our results are affected only slightly by this approximation.} $q^2=(189$ GeV)$^2$.  We note that this expression depends significantly on the renormalization scale $\mu$, which is undetermined in our analysis.\footnote{We note that our expression for the direct-detection cross section, Eq.~(\ref{ddxsect}), is also dependent on the choice of renormalization conditions.  For high-energy processes, however, the small logarithms
are more strongly dependent upon the renormalization conditions.  Here, we quote our results for a specific value of $\mu$.   Additionally, we point out that the 5/3 term appearing in Eqs.~(\ref{eq:eemmttop1}) and (\ref{eq:eemmttop2}) is required to maintain consistency with the renormalization conditions applied in Eq.~(\ref{ddxsect}).   }  To obtain a rough constraint, we take the reference value $\mu=1$ TeV, obtaining 
\begin{align}
\label{eq:eemmttres1}
\frac{-1}{(\mbox{300 GeV})^2}< \sum_{i=1,2}\frac{k_{i\mu\mu}(k_{i\mu\mu}+k_{i\tau\tau})}{M_{X_i}^2}<\frac{1}{(340\mbox{ GeV})^2},\\
\label{eq:eemmttres2}
\frac{-1}{(\mbox{250 GeV})^2}< \sum_{i=1,2}\frac{k_{i\tau\tau}(k_{i\mu\mu}+k_{i\tau\tau})}{M_{X_i}^2}<\frac{1}{(290\mbox{ GeV})^2}.
\end{align}
Varying $\mu$ between 500 GeV and 2 TeV only changes the scales in Eqs.~(\ref{eq:eemmttres1}) and~(\ref{eq:eemmttres2})  by $30-50$ GeV.

We also consider the corrections to the $Z$ couplings to $\mu^+\mu^-$ and $\tau^+\tau^-$, as shown by the diagrams in Fig.~\ref{collider}.  The diagram in Fig.~\ref{zdecay1} is finite when added to external leg corrections; it rescales the $Z$ vertex, giving a new contribution to the $Z$ partial width.\footnote{The diagram in Fig.~\ref{LEPLHC} will also contribute to the $Z$ partial widths.  However, after the application of the limits obtained in Eqs.~(\ref{eq:eemmttres1}) and (\ref{eq:eemmttres2}), we find that the contribution of Fig.~\ref{LEPLHC} to the partial width is subdominant to that of Fig.~\ref{zdecay1}.  For simplicity, we ignore it here. }  The possible new-physics contribution to the partial width of $Z\rightarrow\mu^+\mu^-$ requires
\begin{equation}
\displaystyle\sum_i \frac{k_{i\mu\mu}^2}{M_{X_i}^2} < \frac{1}{(200 \text{ GeV})^2}.
\end{equation}
No significant constraint is obtained from the partial width of $Z\rightarrow\tau^+\tau^-$. 

Additionally, the diagram in Fig.~\ref{LEPLHC} can correct the vector couplings of the $Z$ to muons or taus.  This diagram diverges and leads to kinetic mixing between the $Z$ and $X$ bosons and will depend on the renormalization conditions chosen.  Here, we take $\mu=1$ TeV and consider constraints from the vector coupling of the $Z$ to taus, $g_{V\tau} = -0.0366\pm 0.0010$~\cite{ALEPH:2005ab}.  Taking the 95\% CL range as an approximate measure of the size of a possible new-physics contribution, we obtain
\begin{equation}
\label{eq:gtau}
\left|\displaystyle\sum_{i}\frac{k_{i\tau\tau}(k_{i\mu\mu}+k_{i\tau\tau})}{M_{X_i}^2}\right|<\frac{1}{(330\text{ GeV})^2}.
\end{equation}
On the other hand, we do not get any useful constraint from $g_{V\mu}$.

A few comments must be made on these results.  First, we must stress that, to obtain a rigorous result, it is not adequate to treat corrections to the $Z$ vertex and contributions to effective operators separately; a full fit, simultaneously including all new physics contributions, should be performed.  Additionally, we note that our choice for the renormalization scale, $\mu=1$ TeV, although reasonable, is arbitrary.  For these reasons, our constraints derived from LEP results should be taken as approximate.  We will return to this when we discuss the parameter scans in Section \ref{scans}.

Lastly, if the $X$ were light enough to be produced at LEP, i.e., $M_X\lesssim 200$ GeV, it could be observed via resonance production in the $\mu^+\mu^-$, $\tau^+\tau^-$, or missing energy final states via the diagram in Fig.~\ref{LEPLHC}.  Searches for the resonant peak of a sneutrino decaying to $\mu^+\mu^-$ are relevant for $M_X$ in the range $100<M_X<200$ GeV~\cite{Abbiendi:1999wm}.  However, the cross section for $X$ production is loop-suppressed, and since the sneutrino search is only valid for $X$ widths $<1$ GeV,  only  small, isolated regions in the $M_{X_i}-k_{i\mu\mu}$ parameter space are constrained.  Similarly, analyses constraining invisibly-decaying resonances at LEP are largely insensitive to our scenario~\cite{Fox:2011fx}.

\subsubsection{LHC}

The ATLAS and CMS experiments have searched directly for resonances in the $\mu^+ \mu^-$ final state, placing limits on the cross section of inclusive resonant production of a new particle and subsequent decay into $\mu^+\mu^-$~\cite{Collaboration:2011dca, Chatrchyan:2011wq}.  Lepton-flavored $X$ bosons can be produced on resonance at the LHC, via the diagram in Fig.~\ref{LEPLHC}.   
As was the case with the LEP observables above, this diagram depends on renormalization conditions.  For this reason, we stress that a rigorous prediction for the cross section for $X$ production at LHC is not possible within our model-independent framework.   However, approximate cross sections possible with the LFDM framework can be achieved, and the results from the LHC experiments do not yet seem to significantly constrain the LFDM parameter  space.  We return to this when we discuss the parameter scans in Section~\ref{scans}.

\section{Parameter Scans}
\label{scans}

\begin{figure}[htbp!]
\begin{center}
\includegraphics[width=0.6\textwidth]{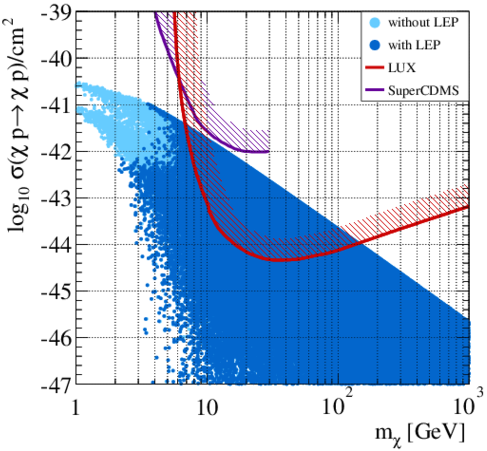}
\caption{The dark- and light-blue points correspond to the scan of the direct-detection cross section, as discussed in Section~\ref{scans}, with and without the LEP constraints, respectively. The red and purple lines are the 90\% CL upper limits from the LUX~\cite{Akerib:2013tjd} and SuperCDMS~\cite{Agnese:2014aze} experiments, which have been rescaled by a factor of $Z^2/A^2$ for Xe and Ge, respectively.  }  
\label{ddscan}
\end{center}
\end{figure}

We illustrate the available phase space for LFDM by performing parameter scans by randomly selecting the values of the couplings and the masses of the gauge bosons in Eq.~(\ref{dim4noLFV1DM}).  We randomly select values of $k_{i\mu\mu}$, $k_{i\tau\tau}$, $k'_{iL}$, $k'_{iR}$, $h_{\mu\tau}$, $h'_L$, and $h'_R$ (where $i=1,2$) within the somewhat-arbitrary range -1 and 1.    The masses of the gauge bosons, $M_{X_1}$, $M_{X_2}$, and $M_Y$ are randomly sampled uniformly between 100 GeV and 10 TeV, and the value of $m_\chi$ is fixed by  requiring that the relic density be the observed value.  As discussed in Section~\ref{indirectdetection}, small to moderate values of $m_\chi$, i.e., between several GeV and ${\cal O}(\mbox{few}\times 10\mbox{ GeV})$, may be disfavored given results from indirect detection.  We do not impose any constraint on the value of $m_\chi$ when performing the parameter scan, since this amounts to only changing the range of the $x$-axis in Fig.~\ref{ddscan}. 

\begin{figure}[htbp!]
\begin{center}
\subfigure[]{\label{mxplot}\includegraphics[width=0.45\textwidth]{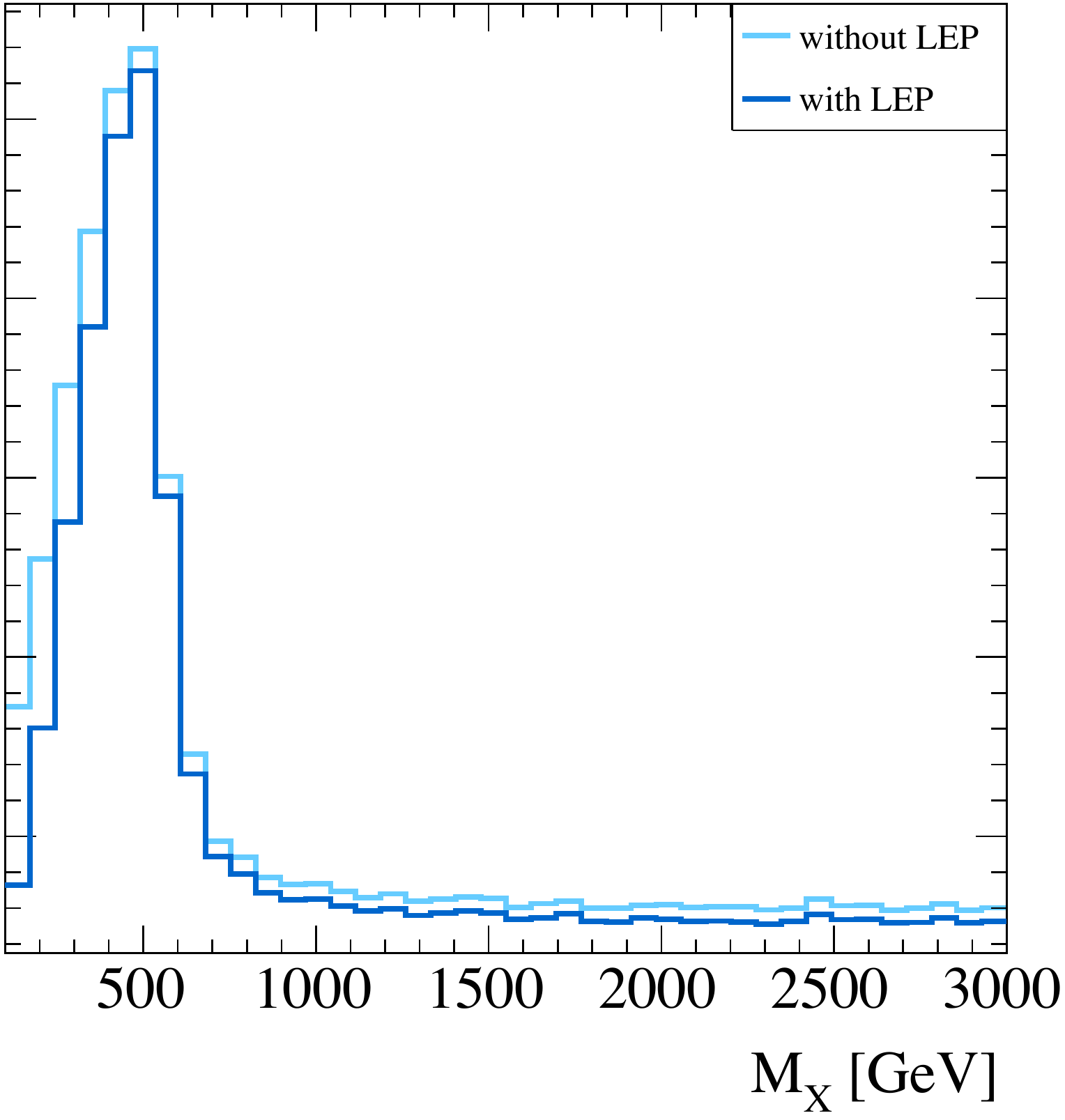}}
\hspace{0.25in}
\subfigure[]{\label{myplot}\includegraphics[width=0.45\textwidth]{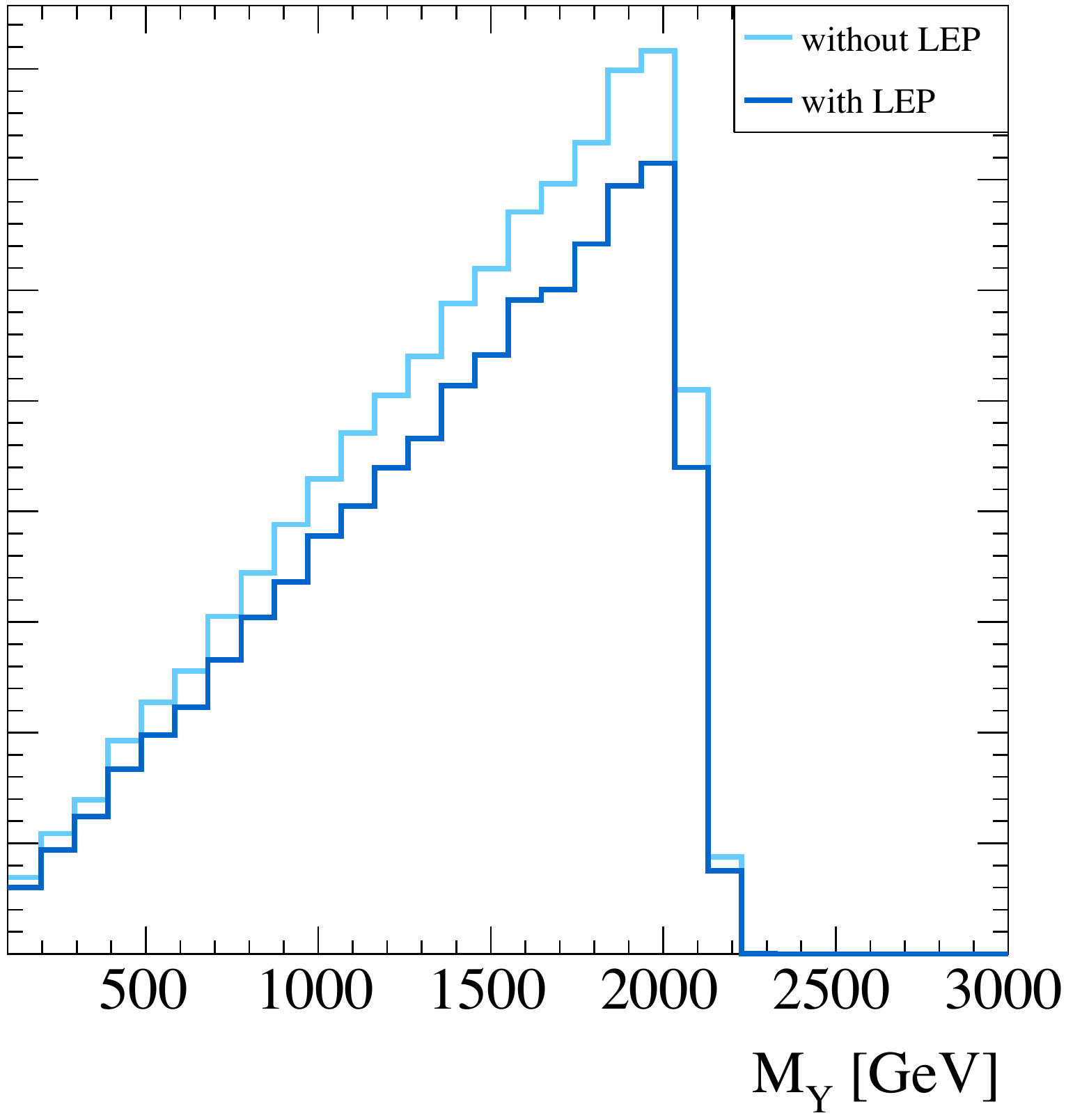}}
\caption{ The same scans to produce Fig.~\ref{ddscan} are plotted here for the distribution of the values of (a) $M_{X_1}$ and $M_{X_2}$ (here combined and called $M_X$) and (b) $M_Y$.  The lighter of the two $X$ gauge bosons is preferred to have a mass of  a few hundred GeV, while the mass of heavier one is relatively unconstrained from above.  The $Y$ gauge boson cannot have a mass $\gtrsim 2$ TeV due to constraints from the muon $g-2$.  The $y$-axes of these plots are not identical and are in arbitrary units.  }  
\label{massplots}
\end{center}
\end{figure}

For each set of couplings and masses, we require that the theoretical and experimental values of the muon $g-2$ are within 95\% CL of each other, while avoiding tension with the measured rates of neutrino trident production and $\tau^-\rightarrow \mu^- \nu_\tau \overline{\nu}_\mu$, as described in Sections~\ref{trident} and~\ref{taudecays}, respectively. For a point in the phase space that passes these requirements, a light-blue point is drawn for the value $m_\chi$ and the direct-detection cross section.  A dark-blue point is drawn for those regions of the phase space that additionally satisfy the constraints from LEP in Eqs.~(\ref{eq:eemmttres1})$-$(\ref{eq:gtau}).  We reiterate that the constraints from LEP are approximate, and the dark-blue points should be taken as illustrative of the rough constraints from LEP.   We compare these scans of the available LFDM phase space with the results from LUX~\cite{Akerib:2013tjd} and SuperCDMS~\cite{Agnese:2014aze}, as shown in Fig.~\ref{ddscan}.  The LEP constraints affect only the light $X$ gauge bosons with masses of only a few hundred GeV or less.    
We randomly sample over $10^9$ points in our phase space, and we find that,  for a wide range of couplings, the lighter of the two $X$ gauge bosons is preferred to have a mass of  a few hundred GeV, while the preferred value of $M_Y$ is multiple TeV.  Histograms of the values of $M_{X_1}$, $M_{X_2}$, and $M_Y$ are shown in Fig.~\ref{massplots}.\footnote{We have checked and found that $M_{X_1}, M_{X_2}, M_Y < m_\chi$ for only $\mathcal{O}$(1\%) of  phase-space points covered in the scans.  This justifies our assumption, made in Section~\ref{directdetection}, that dark-matter annihilations to $X$ and $Y$ gauge bosons can be neglected.}

\begin{figure}[htbp!]
\begin{center}
\includegraphics[width=0.6\textwidth]{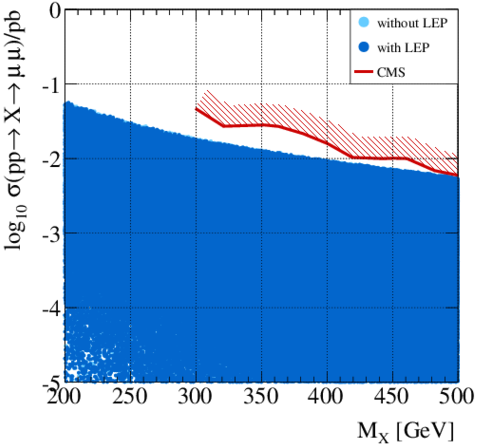}
\caption{The same scans to produce Fig.~\ref{ddscan} are plotted here for the resonant cross section of $pp\rightarrow X \rightarrow \mu^+ \mu^-$ at the LHC, for $\sqrt{s}=7$ TeV and a renormalization scale $\mu=1$ TeV, compared with the 95\% CL upper limits from CMS~\cite{Chatrchyan:2012it}. Interference with the SM has been neglected and can significantly affect the results, as discussed in Section~\ref{scans}.}  
\label{lhcscan}
\end{center}
\end{figure}

We then investigate the implications of LFDM for resonant searches at the LHC.  Scanning over the same points as above, we roughly estimate the $pp\rightarrow X$ cross sections achievable at the LHC, assuming a renormalization scale of $\mu = 1$ TeV and $\sqrt{s} = 7$ TeV.  We use {\sc MadGraph5}~\cite{Alwall:2014hca} to perform leading-order resonant cross-section calculations for production of $pp\rightarrow X \rightarrow\mu^+\mu^-$ arising from the diagram in Fig.~\ref{LEPLHC}.  We do not do a full mixing calculation, and we neglect interference between Fig.~\ref{LEPLHC} and the SM.\footnote{We calculate the resonant cross section for $pp \rightarrow X \rightarrow \mu^+ \mu^-$ for reference values of the $X$-lepton couplings and width $\Gamma_X$. Because the cross section, to a good approximation, scales as the inverse of $\Gamma_X$, we are able to rescale these {\sc MadGraph5} results to account for the varying values of the $X$-fermion couplings in the scans.}  We find that the resulting cross sections in the LFDM scenario for $\mu = 1$ TeV and couplings close to unity are within a factor of three of the current limits from CMS~\cite{Chatrchyan:2012it} for $300\mbox{ GeV}<M_X<500$ GeV, even when constraints from LEP are included.  Our results are shown in Fig.~\ref{lhcscan}.

However, we note that these results are only approximate.   In particular, the cross section for $X$ production becomes small as $M_X$ approaches the renormalization scale $\mu$; in this case, the approximation of neglecting the effects of SM interference with the diagram of Fig.~\ref{LEPLHC} becomes invalid.  We find that the effects of interference with the SM are significant and cannot be neglected for $300\mbox{ GeV}<M_X<500$ GeV.  Additionally, interference with the SM can cause the shape of the resonance to deviate from that of a Breit-Wigner, which can  affect the sensitivity of searches for narrow resonances.  Due to these effects, models with low renormalization scales\footnote{We note that the renormalization scale $\mu$ depends on the ultraviolet completion of a particular model, and is undetermined in our analysis.} might require dedicated experimental analyses.  For these reasons, we extend the $x$-axis of Fig.~\ref{lhcscan} to only $M_X = 500$ GeV.  However, as the amplitude resulting from Fig.~\ref{LEPLHC} grows in size with increasing $\mu$, our results do indicate that models with somewhat larger renormalization scales would be less affected by interference with the SM and would give cross sections similar to those shown in Fig.~\ref{lhcscan}.  Therefore, we point out that if a local excess is seen at the LHC in the $\mu^+\mu^-$ channel, but not in the $e^+e^-$ channel, LFDM would be a candidate explanation. 

To illustrate the interplay between the observables in the parameter scans, we briefly mention two toy models.  First, we consider a $U(1)$ flavor model where the $\mu$ and $\tau$ flavors have equal and opposite couplings to a single $X$ boson (this is the well-known $U(1)$ $L_\mu - L_\tau$ model, the discussion of such a setup can be found in Ref.~\cite{Heeck:2011wj} and references therein).  We allow mixing between flavored and unflavored states in the dark sector, and thus take the magnitude of the right- and left-handed couplings of the dark matter to the $X$ equal to or less than that of the muon.  We find that such a model can satisfy the relic density, direct detection, $g-2$ and tau-decay constraints; $X$ production at colliders is severely suppressed by cancellation between $\mu$- and $\tau$-loop contributions.  However, this toy model is ruled out due to tension between $g-2$ and the limits from neutrino trident production.  Second, we briefly consider an $SU(2)$ model, where $\mu$- and $\tau$-flavored fields are placed into a flavor doublet. (For previous use of $SU(2)$ in flavor models, see Ref.~\cite{Heeck:2011wj}.)  This model contains $X$ and $Y$ bosons of equal mass, both of which contribute to $g-2$.  Due to the tension between tau decays, which require the new physics scale for $Y$-mediated processes to be above $\sim 2$ TeV, and $g-2$, which favors a scale below 1.9 TeV, this model is marginally ruled out.  (While the $X$ boson also contributes to $g-2$, this contribution is inadequate to resolve this tension.)  Although it is reasonable to speculate that this latter model could be embedded into a larger model with additional contributions to $g-2$ from $X$ bosons or which contains a mass splitting between the $X$ and $Y$, we do not attempt further model-building here.

Lastly, we briefly mention that the phase space ruled out by direct detection has no model-independent ramification on the discovery potential at the LHC, i.e., the points within the LUX limits in Fig.~\ref{ddscan} inhabit the entire phase space for the LHC.  Any specific mapping between direct detection and hadron colliders must be done with a specific model.  
Additionally, to get a rough idea of the effect indirect detection constraints can have on our LHC results, we repeat the scan in Fig.~\ref{lhcscan} but requiring $m_{\chi}>50$ GeV.  We find that if indirect detection rules out low-mass dark matter candidates, it does not qualitatively change the available phase space for LFDM at the LHC.


\section{Lepton-Flavor Violation}
\label{LFV}

Up to this point, we have only considered the case where LFV among charged leptons is negligible.  Here, we briefly explore the possibility of allowing a small amount of $\mu-\tau$ flavor violation, still assuming that the flavor gauge bosons couple negligibly to electrons.  The interactions introduced in this section, if taken in isolation, are likely inadequate to account for muon $g-2$.  However, we present them as an indication of the level of LFV which may be allowed in more complete models of flavored dark matter.

The notation using $X$ and $Y$ bosons in Eq.~(\ref{dim4noLFV1DM}) is no longer adequate if we allow for LFV among charged leptons, so we will discuss LFV using effective four-lepton operators
\begin{equation}
\label{eq:fveffop}
{\cal O}_{ijkl} \equiv \frac{C_{ijkl}}{ (1+\delta) \Lambda^2}\left(\overline{\ell}_i \Gamma^{\alpha}_{ij} \ell_j\right) \left(\overline{\ell}_k \Gamma_{\alpha kl} \ell_l\right), 
\end{equation}  
where $\ell_a$ is a charged lepton of either muon $(a=\mu)$ or tau $(a=\tau)$ flavor, and $\delta=1(0)$ if $i=k$ and $j=l$ ($i\neq k$ or $j\neq l$).  While we assume vector interactions in the flavor interaction basis, we have generalized the Lorentz structure from $\gamma^{\alpha}$ to $\Gamma^{\alpha}$, since the interactions may deviate from purely vector when rotating the right- and left-handed sectors to the mass basis.

Only three of these operators, ${\cal O}_{\tau\mu\mu\mu}$, ${\cal O}_{\mu\tau\tau\tau}$, and ${\cal O}_{\tau\mu\tau\mu}$, violate lepton flavor.  The first of these, ${\cal O}_{\tau\mu\mu\mu}$, will contribute to the decay $\tau^-\rightarrow\mu^-\mu^-\mu^+$, which is constrained to have a branching fraction of less than $2.1\times 10^{-8}$ at $90\%$ CL~\cite{Agashe:2014kda}.  The operator ${\cal O}_{\mu\tau\tau\tau}$ can be constrained by assuming that it must be consistent with the branching fraction of $\tau^-\rightarrow\mu^-\gamma$, which has a value less than $4.4\times 10^{-8}$ at $90\%$ CL~\cite{Agashe:2014kda}.  Lastly, because of isospin symmetry, one generically expects that ${\cal O}_{\tau\mu\tau\mu}$ can give rise to the process $\tau^-\rightarrow\mu^-\nu_{\mu}\bar{\nu}_{\tau}$.  Because
this process is experimentally indistinguishable from the SM process $\tau^- \rightarrow \mu^- \nu_\tau \overline{\nu}_\mu $, a constraint on the branching fraction for this process can be obtained by assuming that it must be consistent with the measured ratio of $\Gamma(\tau^-\rightarrow\mu^-\nu_{\tau}\bar{\nu}_{\mu})/\Gamma(\tau^-\rightarrow e^-\nu_{\tau}\bar{\nu}_{e})$ at 90\% CL~\cite{Agashe:2014kda}.  The limits on $|C_{ijkl}|/\Lambda^2$ are shown in Table \ref{table:LFV}.  For simplicity, we only consider the cases $\Gamma^\alpha_{ij}=\Gamma^\alpha_{kl}=\gamma^\alpha,\gamma^\alpha(1\pm\gamma_5)/2$; in general, $\Gamma^\alpha_{ij}$ need not equal $\Gamma^\alpha_{kl}$.  
\begin{table}[ht]
\centering
\begin{tabular}{|c|c|c|}
\hline
$|C_{ijkl}|/\Lambda^2$ & $\Gamma^\alpha_{ij}=\Gamma^\alpha_{kl}=\gamma^\alpha$ & $\Gamma^\alpha_{ij}=\Gamma^\alpha_{kl}=\gamma^\alpha(1\pm\gamma_5)/2$\\
\hline
$|C_{\tau\mu\mu\mu}|/\Lambda^2$ & $1/(15\mbox{ TeV})^2$ & $1/(11\mbox{ TeV})^2$\\
$|C_{\mu\tau\tau\tau}|/\Lambda^2$ & $1/(2.5\mbox{ TeV})^2$ & $1/(1.5\mbox{ TeV})^2$ \\
$|C_{\tau\mu\tau\mu}|/\Lambda^2$ & $1/(0.7\text{ TeV})^2$ & $1/(0.5\text{ TeV})^2$ \\
\hline
\end{tabular}
\caption{Limits on $|C_{ijkl}|/\Lambda^2$, for the cases $\Gamma^\alpha_{ij}=\Gamma^\alpha_{kl}=\gamma^\alpha,\gamma^\alpha(1\pm\gamma_5)/2$ at 90\% CL. }
\label{table:LFV}
\end{table}

As shown in Table~\ref{table:LFV}, the operator ${\cal O}_{\tau\mu\mu\mu}$ is more strongly constrained than the other two flavor-violating operators.  This implies, for example, that a flavor gauge boson which couples to both $\bar{\mu}\mu$ and $\bar{\tau}\mu$ must have a mass $> {\cal O}(10 \mbox{ TeV})$ or at least one somewhat small coupling.  
However, these constraints do leave open another interesting possibility.
We can introduce a $U(1)_\text{LFV}$ flavor gauge boson $V$ which interacts with a single generation of leptons.  We take that generation to be only quasi-aligned with the charged lepton mass eigenstates. Specifically, we take the $V$ boson to interact with an admixture of $\mu$ and $\tau$, 
\begin{equation}
\label{dim4LFV}
\mathcal{L}_\text{LFV} \supset  k_{3}V_\alpha(\overline{L}_3\gamma^\alpha L_3  + \overline{\ell}_{R3}\gamma^\alpha \ell_{R3} + \overline{\nu}_{R3}\gamma^\alpha \nu_{R3}), 
\end{equation}
where
\begin{eqnarray}
\label{LFV1}L_3 &=& L_\tau\cos\theta_L + L_\mu\sin\theta_L, \\
\label{LFV2}\ell_{R3} &=& \tau_R \cos\theta_{R} + \mu_R \sin\theta_{R}, 
\end{eqnarray}
and the $\tau$ component dominates, i.e., $\cos\theta_L \sim\cos\theta_{R}\sim 1$. For simplicity, we ignore neutrino mixing and neglect a possible complex phase in Eqs.~(\ref{LFV1}) and (\ref{LFV2}) and consider the case $\sin\theta_L\sim\sin\theta_{R}\equiv\sin\theta$, although this does not hold generally.  In this case, ${\cal O}_{\tau\mu\mu\mu}$ would be suppressed by three powers of $\sin\theta$, while ${\cal O}_{\tau\mu\tau\mu}$ and contributions to $g-2$ would be suppressed by two such factors.  Additionally, a four-$\tau$ operator, ${\cal O}_{\tau\tau\tau\tau}$,  suffers no such suppression,  and is experimentally unconstrained.   

We note that these interactions which violate charged-lepton flavor, taken alone, are not useful to account for the discrepancy in $g-2$.  Presumably they would need to be incorporated into a more complete model of LFDM.  A $V$ boson with an electroweak-scale mass, i.e., $\sim 1$ TeV, and a mixing angle as large as $\sin\theta\sim {\cal O}(0.1)$ is phenomenologically allowed.  Thus, it is possible to have significant TeV-scale $\mu-\tau$ flavor violation, as long as the couplings are chosen carefully. However, we do not attempt to incorporate flavor violation into a larger model here.

\section{Summary and Conclusions}
\label{conclusion}

Motivated by the apparent tension between the dark-matter relic density and null results from direct detection, as well as the 3.6$\sigma$ discrepancy between the measured and predicted value of the muon $g-2$, we posit that dark matter and the mystery of lepton flavor may be related.  In our scenario, dark matter does not merely couple primarily to leptons, but it shares a common gauged flavor interaction with the leptons in the SM, under which only the muon and tau families are charged.   
We investigate this scenario by creating a Lagrangian composed of $d=4$ operators, allowing the SM leptons to have both flavor-diagonal and flavor-changing couplings to flavor gauge bosons.  Considering the muon $g-2$ and the lack of observed LFV among charged leptons, we take such interactions to be purely vector in the charged lepton sector.  However, flavor mixing may be large in the dark sector, so different couplings for left- and right-handed dark matter are permitted.   We then constrain our scenario using the dark-matter relic density and direct detection, low-energy measurements, constraints from LEP, and neutrino trident production, while demanding consistency with the measured value of the muon $g-2$.  Future prospects for LHC and direct detection are investigated.

In the LFDM scenario, both the dark-matter relic density and the muon $g-2$ suggest that such interactions between dark matter and leptons can exist at the electroweak scale, i.e., $\mathcal{O}$(100 GeV) - $\mathcal{O}$(1 TeV).  
While direct detection, $\tau^-\rightarrow \mu^- \nu_\tau \overline{\nu}_\mu$, and $\mu^+\mu^-$ and $\tau^+\tau^-$ production at LEP do constrain the couplings and masses of the flavor gauge bosons, significant parameter space remains.  
This is illustrated in Figs.~\ref{ddscan} and~\ref{lhcscan}, where we show the possible values of the cross sections for direct detection and resonant production at the LHC, allowed by present constraints.  We find that there are large ranges of flavor gauge boson masses and couplings which give cross sections below but near current sensitivity for both direct detection experiments and resonance searches at the LHC.

Future measurements bring significant hope for further investigations of LFDM.  The particular parameterization of LFDM explored in this work can account for all experimental observations, including the muon $g-2$, if there are both flavor-diagonal interactions at a scale of a few hundred GeV and off-diagonal interactions at a few TeV.  
Improvements in the sensitivity of direct-detection experiments, the precision of the muon $g-2$ measurement, the precision of the branching fraction of $\tau^-\rightarrow \mu^- \nu\overline{\nu}$, the precision of the rate of neutrino trident production, and resonance searches at the LHC can further constrain, or potentially discover, LFDM.   Additionally, flavor-violating observables, such as $\tau^-\rightarrow\mu^-\mu^-\mu^+$, could be useful in constraining or confirming specific models of LFDM, while processes that involve electrons, e.g., $\mu\rightarrow 3e$, $\mu\rightarrow e\gamma$, and $\mu-e$ conversion, are expected to be highly suppressed.

The particular parameterization of LFDM considered in this work permits two contributions to the value of the muon $g-2$:~one contribution from an $X$ boson and another from a $Y$ boson.  Taken separately, these two interactions are nominally ruled out via neutrino trident production and tau decays, respectively. If contributions from $X$ and $Y$ bosons are considered simultaneously, then the constraints from the muon $g-2$, neutrino trident production, and tau decays can each be satisfied at 95\% CL, though there remains some overall tension between this parameterization of LFDM and the data.  We note that the theory and experimental values of the muon $g-2$ are poised to change in the near future, and the LFDM framework can accommodate a range of contributions to $g-2$, particularly those somewhat smaller than the value currently suggested by experiment. If, for example, the muon $g-2$ agrees with the SM prediction, higher-mass $X$ and $Y$ bosons would easily account for the data.  As such,  $g-2$ is not necessary for LFDM, though we suggest that given our present framework, we might expect future measurements and future theory calculations of $g-2$ to come into better agreement.

We also point out a few possible extensions to this work.  While we have assumed negligible couplings of the flavor gauge bosons to electrons, the addition of such small couplings could be investigated.  Additionally, the possibility of charged-lepton flavor violation, only touched upon here, could be studied in more detail.  And, of course, the construction of more concrete models of LFDM would also be a worthy endeavor.  

In conclusion, the LFDM framework can address the dark-matter relic density and constraints from LEP, while explaining the lack of an observed signal in direct-detection experiments and the muon $g-2$.  Lastly, it holds significant promise for future experiments.

\begin{acknowledgments}
We thank Andr\'{e} de Gouv\^{e}a for providing useful feedback and comments and Julian Heeck for pointing out the constraints from neutrino trident production.  The work of AK is sponsored in part by the DOE grant No.~DE-FG02-91ER40684 and by the Department of Energy Office of Science Graduate Fellowship Program (DOE SCGF), made possible in part by  the American Recovery and Reinvestment Act of 2009, administered by ORISE-ORAU under contract no.~DE-AC05-06OR23100.  JK is supported in part by the DOE grant No.~DE-FG02-97ER41029.   The work of AS is supported in part by the DOE contract No.~DE-AC02-98CH10886 (BNL).
\end{acknowledgments}

\bibliography{bib}{}

\end{document}